\documentstyle[twocolumn,prc,aps]{revtex}
\input epsf
\newcommand{\be}{\begin{eqnarray}}
\newcommand{\ee}{\end{eqnarray}}
\begin{document}

\twocolumn[\hsize\textwidth\columnwidth\hsize\csname @twocolumnfalse\endcsname
\title{Diquark Bose Condensates in High Density Matter and Instantons}
\author{R.~Rapp$^1$, T.~Sch{\"a}fer$^2$, 
        E.V.~Shuryak$^1$ and M.~Velkovsky$^3$}
\address{
$^1$ Department of Physics and Astronomy, State University of New York, 
     Stony Brook, NY 11794-3800\\
$^2$ Institute for Nuclear Theory, Department of Physics, University of 
     Washington, Seattle, WA 98195, USA\\ 
$^3$ Nuclear Theory Group, Brookhaven National Laboratory, Upton, 
     NY 11973-5000}
\date{\today}
\maketitle    

\begin{abstract} 
 Instantons lead to strong correlations between up and down quarks 
with spin zero and anti-symmetric color wave functions. In cold 
and dense matter, $n_b>n_c\simeq 1 fm^{-3}$ and $T<T_c\sim$ 50 MeV,
these pairs Bose-condense, replacing the usual $\langle \bar qq 
\rangle$ condensate and restoring chiral symmetry. At high density,
the ground state is a color superconductor in which diquarks
play the role of Cooper pairs. An interesting toy model is provided
by QCD with two colors: it has a particle-anti-particle symmetry 
which relates $\langle\bar qq\rangle$ and $\langle qq\rangle$ 
condensates.
\end{abstract}  
\vspace{0.1in}
]
\begin{narrowtext}   
\newpage

 The properties of hadronic matter under extreme conditions are 
subject to intense theoretical studies, numerical simulations on 
the lattice, and experimental efforts using high energy heavy-ion 
collisions. Substantial progress has been made with respect to  
high {\em temperature} QCD matter, while the (more difficult) problem 
of cold {\em dense} matter is much less understood. This is partly
due to the fact that up to now, lattice simulations have not been 
able to overcome the numerical problems associated with the fact
that at non-zero chemical potential the fermionic determinant is
complex (see~\cite{Barbour} for a recent review). This is unfortunate,
because the phase structure of dense matter may be very rich. Several
intermediate phases between nuclear and quark matter have been 
proposed, e.g. states with mesonic (pion or kaon~\cite{mesonic_cond}) 
condensates. 

  Due to the asymptotic freedom of QCD one expects very dense matter 
to resemble an ideal Fermi gas of quarks. The system is quite similar
to a cool electron plasma, with Debye-screening of color fields at 
momentum scales $p<M_D\sim g\mu$, collective plasmon excitations, 
etc.~\cite{Shu_78}. Since the Coulomb interaction between quarks 
of {\em different colors} is attractive, it was realized early on 
that the quark plasma should be a superconductor, due to the formation 
of Cooper pairs on the Fermi surface~\cite{BL}. The magnitude of the 
corresponding gap $\Delta$ and the critical temperature $T_c$ were 
estimated to be in the MeV range. 

  In this article we show that non-perturbative effects lead to  
diquark condensates with $\Delta,T_c$ about two orders of magnitude 
larger. These condensates are generated by the instanton-induced
interactions between light quarks~\cite{tHooft}. For two flavors
(up and down) the $(\bar qq)$ interaction is 
\be 
\label{l_mes}
{\cal L} &=& G{1\over 8 N_c^2}
 \left[(\bar\psi \tau^- \psi)^2+
 (\bar\psi \tau^- \gamma_5 \psi)^2 \right],
\ee
where we have added the interaction in the direct and exchange 
channels and dropped color octet terms. $N_c$ is the number of 
colors and $\tau^-=(\vec\tau,i)$ is an isospin matrix. We will 
specify the coupling constant $G$ below. There is pervasive 
evidence for the importance of this interaction from (i) 
phenomenological studies of current correlation functions in 
QCD, (ii) the success of hadronic spectroscopy in the instanton 
liquid model, and (iii) studies of instantons and their effects 
on the lattice, see \cite{SS_97} for a review of these issues. 

  The result (\ref{l_mes}) can be Fierz-rearranged into a $(qq)$ 
interaction. One obtains  
\be 
\label{l_diq}
{\cal L} &=& 
G \left\{
 -{1\over 16 N_c (N_c-1)}
 \left[ (\psi^T C \tau_2 \lambda_A^n \psi)
        (\bar\psi\tau_2 \lambda_A^n C \bar\psi^T) 
  \right.\right.\\
  & &  \left. \hspace{2.0cm}\mbox{}
       +(\psi^T C \tau_2 \lambda_A^n \gamma_5 \psi)
        (\bar\psi \tau_2 \lambda_A^n \gamma_5 C \bar\psi^T) \right]  
               \nonumber\\
  & &   \left. \mbox{}
      \! +{1\over 32 N_c (N_c+1)}
        (\psi^T C \tau_2 \lambda_S^n \sigma_{\mu \nu} \psi)
        (\bar\psi \tau_2 \lambda_S^n \sigma_{\mu \nu} C \bar\psi^T) 
        \right\} \nonumber
\ee 
Here, $C$ is the charge conjugation matrix, $\tau_2$ is the anti-symmetric 
Pauli matrix, $\lambda_{A,S}$ are the anti-symmetric (color $\bar 3$) and 
symmetric (color 6) color generators. The effective lagrangian (\ref{l_diq}) 
provides a strong attractive interaction between an up and a down quark with 
anti-parallel spins ($J^{P}=0^{+}$) in the color anti-triplet channel, and
a repulsive interaction in the $0^-$ channel. $0^+$ quark pairs couple to 
the diquark current $S^a_{dq} = \epsilon_{abc} u^T_b C \gamma_5 d_c$.
Phenomenological implications of the instanton-induced interaction in 
this channel were first discussed in connection with spin-dependent 
forces in baryons \cite{SR}, challenging the conventional wisdom that 
spin splittings are due to one-gluon exchange. The importance of diquark 
degrees of freedom is perhaps most obvious in baryons that contain one 
very heavy quark. Quantitative studies of instanton effects 
in baryon spectroscopy (both light and heavy-light systems) were done 
in~\cite{SSV}. The conclusion was that the instanton effects encoded
in (\ref{l_diq}) are indeed strong enough to reproduce the observed 
spin-splittings, and that the nucleon has a very large overlap with 
the current $\epsilon_{abc}(u^T_aC\gamma_5d^b)u^c=S^a_{dq}u^a$.
 Since there is 
no confinement in the instanton model, one can compare the diquark mass
to the two constituent quark threshold. The result is a deeply bound
scalar diquark $2m_q-m_{Sdq}\simeq 200-300$ MeV~\cite{note1}, whereas
all other channels (vectors and axial-vectors, color 6 diquarks, etc.)
are at most weakly bound.

  The possible role of diquark clusters in quark matter was discussed 
in~\cite{DS}. It was noted that a loosely bound ``third'' quark in the 
nucleon may find a partner in dense matter. However, as we show below, 
this effect is less important than Bose condensation. In general, even 
if we focus only on the instanton-induced interaction, the situation is 
quite involved due to the competing attraction in the $\bar qq$ channel. 
At small density, the attraction in the $\bar qq$ channel is stronger
than the one in the $qq$ channel (compare (\ref{l_mes}) with (\ref{l_diq})), 
$\bar qq$ pairs condense and chiral symmetry is broken. At large density, 
Pauli-blocking suppresses the $\bar qq$ interaction, but the $qq$ 
interaction leads to an
instability near the Fermi surface. The mechanisms for $qq$ and $\bar qq$
condensation are quite similar, both are based on the formation of a   
gap in the fermion spectrum at the surface of the Dirac/Fermi sea, 
respectively. We will defer an attack on the coupled problem to a
separate publication, and in this work only consider the low and high 
density parts separately. 

 At low baryon density, we ignore modifications of the vacuum and keep 
$\langle \bar q q \rangle$ fixed. In this limit instanton-based models 
realize the minimum of the energy per quark in scalar diquarks, not in 
nucleons. Only by taking confinement into account, one finds that
at low density the ordinary nuclear 
matter is favored over a diquark Bose gas. At high densities we ignore 
interactions with anti-quarks,  assume a deconfined chirally symmetric 
quark matter, and calculate how instantons operate at the Fermi surface. 
For pedagogical reasons, let us first discuss the special case of QCD 
with two colors $N_c=2$.  

2. In the massless case $N_c$=2 QCD has an additional Pauli-G{\"u}rsey 
symmetry (PGSY)~\cite{PG,DP_diquarks}, which mixes quarks with anti-quarks. 
As a result, diquarks (=baryons) are color singlet states that are degenerate 
with the corresponding mesons. Chiral symmetry breaking then implies 
that some diquarks are Goldstone bosons, that means their mass vanishes
in the chiral limit ($m_q=0$). Their properties follow directly from 
the PGSY. The total number of Goldstone modes is \cite{SV,DP_note,Peskin} 
$2N_f^2-N_f-1$. For $N_f$=1 there is no Goldstone boson
(the $U(1)$ symmetry is anomalous). For $N_c=N_f=2$ one finds five: 
3 pions, the scalar diquark $S$ and its anti-particle $\bar S$. There
is a nice continuity in going from $N_c=2$ to 3: the scalar diquark 
goes from being massless to a deeply bound state! 

  The coset of the full group over the unbroken one is $K=SU(4)/Sp(4)
=SO(6)/SO(5)=S^5$, {\it i.e.}~the effective chiral Lagrangian
describing the Goldstone modes is a sigma model, in which the usual
``chiral circle'' is replaced by a 5-dimensional sphere. The ordinary
vacuum of the massless $N_c=N_f=2$ theory with a non-zero ``mesonic'' 
condensate $\langle \bar q q \rangle$ can simply be rotated into states 
with finite diquark condensate. This rotation costs no energy, so the 
chemical potential $\mu=\partial E/\partial n_q=0$. At the point where
the $\langle qq\rangle$ condensate is as large as the original $\langle
\bar qq\rangle$ condensate, chiral symmetry is restored, but there are 
still 5 Goldstone modes, which are now 3 pions, sigma and $\bar S$.

 For non-zero quark mass the qualitative picture can be understood 
using the corresponding linear sigma model. The potential
\begin{equation} 
 V = \lambda \left( \vec\pi^2 \!+\! \sigma^2 \!+\! S^2 \!+\! \bar S^2
     \!-\! v^2 \right)^2 - A\sigma - \mu^2 \left( S^2 \!+\! \bar S^2 \right)
\end{equation} 
includes the diquark chemical potential $\mu$ and the chirally asymmetric 
mass term $A$. At $\mu=0$ the Goldstone masses are $m_g^2=A/v$, and 
$m_\sigma^2=8\lambda v^2$. For non-zero $\mu$ we can determine the 
$\langle\bar qq\rangle$ and $\langle qq\rangle$ condensates 
$\langle\sigma\rangle$ and $\langle S\rangle$ using the mean 
field approximation. We find
\begin{equation} 
 4\lambda \langle S \rangle \left(\langle \sigma\rangle^2+
\langle S \rangle^2-v^2\right) =2\mu^2 \langle S \rangle \ .  
\end{equation} 
Below the critical chemical potential $\mu_c\simeq m_g/\sqrt{2}$, 
$\langle\sigma\rangle$ is constant and $\langle S\rangle=0$. Above
$\mu_c$, $\langle S\rangle$ increases as 
\begin{equation} 
 \langle S \rangle^2 = {\mu^2\over 2\lambda} + v^2 - {A^2 \over 4 \mu^2} 
\end{equation}
and $\langle \sigma \rangle = m^2_g v/(2\mu^2)$. The energy density is 
$\epsilon =-\mu^2v^2-{3 A^2/(4\mu)^2}$, compared to $\epsilon =-m_g^2
v^2+m_g^4/(16 \lambda)$ for the normal vacuum. 

  Unlike real QCD, $N_c=2$ gauge theory is straightforward to simulate 
on the lattice, since the fermion determinant remains real for $\mu\neq 0$. 
With the exception of some early work using small lattices and the strong
coupling expansion \cite{strongcoupling}, few studies have taken 
advantage of this. Numerical studies of the instanton model for $N_c=2$
at finite density~\cite{S_97} are consistent with the scenario found
above, at large density the $\langle\bar qq\rangle$ condensate is 
replaced by a $\langle qq\rangle$ condensate.

3. Let us return to the low density limit of real QCD with three colors, 
ignoring strangeness. For definiteness, we consider neutron ($udd$) matter 
relevant for stars\cite{note_collisions} in which $ud$ diquarks and $d$ 
quarks compensate each others color and electric charges.
Since scalar diquarks are color anti-triplets, a Bose condensate will
select a direction in color space. If we label this direction red ($k$=3),
our $ud$ diquarks are made of blue and green ($k$=1,2) quarks only.
Their properties are strongly modified by the diquark condensate, 
while the third quark-type (red $d$) is basically unaffected.

  Diquarks are favored over a quark Fermi gas due to both their binding
$m_S<2m_{eff}$ as well as their Bose character. However, it is erroneous 
to conclude that an infinite number of diquarks condenses in the $p=0$ 
state: diquarks are composite objects and, like nucleons, their interaction 
should have a repulsive core. We account for this by introducing a 
scattering length~\cite{note3} $a\simeq 0.3$ fm into the expression 
for the energy per quark in the diquark Bose gas:  
\begin{equation}
{\epsilon^B \over n_q}={4\pi a n_S \over m} 
\left( 1+ {128 \over  15\pi^{1/2}}(a^3 n_S)^{1/2}\right)
\end{equation}
where $n_S$ is the $S$ diquark density. The first term is the mean 
field result, and the second term comes from non-condensed 
diquarks~\cite{LY}. The repulsion makes the pure diquark gas less 
favorable than the optimal Bose-Fermi mixture. 
Results of our calculations are 
shown in  Fig.\ref{fig1_fb}, in which matter consists of (i) a Bose 
gas of $S$ diquarks in chemical equilibrium with (ii) a Fermi gas of 
blue/green quarks, and neutralized in color and electric charge
by an appropriate amount of (iii) red quarks. For definiteness, we 
use the di-/quark masses of 500~MeV and  400~MeV, respectively. 

 To account for confinement we add color strings to our picture. 
For light hadrons, lattice calculations suggest as potential of the
form $V(R)=K |R-R_0| \theta(R-R_0)$ with the usual string tension 
$K\simeq1$ GeV/fm but smoothed out at the origin, here represented
by $R_0\simeq 0.7$ fm (see example in~\cite{FIMT_96}). We evaluate 
the average energy of a string using this schematic potential with  
$R=n_{tot}^{-1/3}$ where $n_{tot}$ is the total density of all 
quarks and diquarks. As can be seen from Fig.\ref{fig1_fb}, strings 
preserve nuclear matter at low density. A diquark/quark/string mixture
has a shallow (meta-stable) minimum leading to a mixed phase at
$n\approx$ 0.5--1 fm$^{-3}$. It
 may be an artefact of the crude model: but  most recent 
lattice results~\cite{Barbour} have some indications for such behaviour.
Note a significant gain in total energy relative to quark 
matter (upper grey curve).  The critical temperature for Bose 
condensation  can be roughly 
estimated from Einstein's ideal gas expression, 
$T_c=3.31 n_{dq}^{2/3}/m$, which is about 100 MeV at the crossing 
point.
\begin{figure}[h]
\epsfxsize=3.0in
\centerline{\epsffile{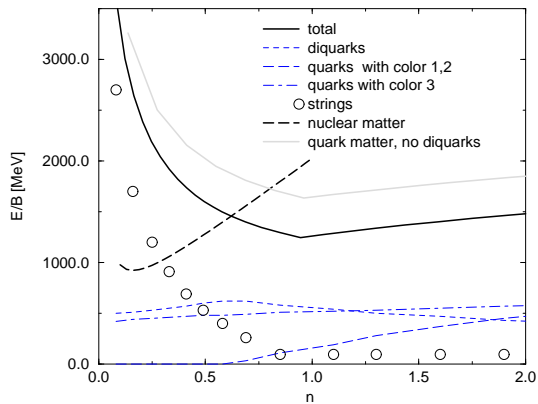}}
\caption[]{ Energy per baryon (in MeV) versus  baryon charge density $n_b$
 (fm$^{-3}$).  }
\label{fig1_fb}
\end{figure}

4. We now turn to the high density limit, with most of the quarks  
forming a Fermi gas  and instanton-induced forces operating only 
near the Fermi surface. The width of this zone (the analogue of 
the Debye frequency in a phonon superconductor) is governed by 
the instanton form factor. 
The fermion zero modes that lead to the 
interaction (2) and determine the form factor  
depend on density~\cite{Abrikosov}. They 
oscillate in the spatial direction, reflecting the presence of
a Fermi surface, and are long range in time direction. As a result, 
the diagrams in the $qq$ channel are infrared divergent, making
the Fermi surface unstable and leading to the formation of a gap.\\ 
The full problem of evaluating the size of the gap is quite involved.  
Instead, we will consider a BCS-type gap equation  where the form factor
is represented by a fixed cutoff $\lambda$$\simeq$0.3~GeV. 
We do not include interactions with anti-quarks explicitly, but 
parametrize the $\mu$-dependence of $\langle \bar q q\rangle$ and the 
constituent quark masses $m_q$ by the factor 
$(\mu_c^3- \mu^3)/\mu_c^3$ for $\mu<\mu_c\simeq$ 500~MeV 
(resembling a linear decrease in density), where the precise value of
the critical chemical potential for chiral restoration, $\mu_c$, will 
be determined in future work.
The gap equation in the most attractive channel
$u_id_j\epsilon_{ij3}$ then reads  
\begin{equation}
1=\frac{8}{(2\pi)^2} \ g_{eff}(\mu) \int_{p_F-\lambda}^{p_F+\lambda} p^2 dp 
\frac{\tanh(\epsilon_p(\Delta)/2T)}{\epsilon_{p}(\Delta)} 
\label{gap}
\end{equation}   
with
$\epsilon_p(\Delta)=[(\omega_p-\mu)^2+\Delta(\mu,T)^2]^\frac{1}{2}$,
$\omega_p^2=p^2+m_q^2$ and $p_F^2=\mu^2-m_q^2$.
For $\mu\ge\mu_c$ the Debye screening of the instanton fields 
becomes effective~\cite{Shu_82}, 
which we account for in the coupling constant 
of the effective fermionic interaction: 
\begin{eqnarray}
g_{eff}(\mu)=C \ (8\pi^2/g^2)^6\int d\rho \ \rho \  
\exp[-8\pi^2/g(\rho)^2] 
  \nonumber \\
\times  \exp[-N_f\rho^2(\mu^2-\mu_c^2)\theta(\mu-\mu_c)]  \  
\exp[-A \rho^2] \ .
\label{geff}
\end{eqnarray} 
The last exponential factor is included to provide a cutoff at large $\rho$.
Using the gap equation in the quark-antiquark channel we fit the value of 
the constant $A$ in vacuum ($T$=$\mu$=0) to a constituent quark mass 
$m_q\simeq400$ MeV. 
In QCD with 3 flavors $g_{eff}(\mu)$ for
 $ud$ quarks is additionally reduced by the factor 
$(m_s^0+m^*_s(\mu))/(m_s^0+m^*_s(0))$ due to the decreasing 
$\langle \bar ss\rangle$-related contribution to $m_s$ 
($m_s^*(0)$$\simeq$200~MeV).
Fig.~2 shows the gap $\Delta(\mu,T=0)$ with $N_f$=3
 for two different  values of the critical chemical potential.
 At large $\mu$ the gap is strongly suppressed by screening effects, 
while at small $\mu$ it is reduced due to the decrease of the 
density of states  at the Fermi surface  
(here the  approach discussed in sect.~3 is more appropriate).
 The maximum gap is approximately 
linear in the critical density and reaches 50-100~MeV.
\begin{figure}[h]
\epsfxsize=3.0in
\centerline{\epsffile{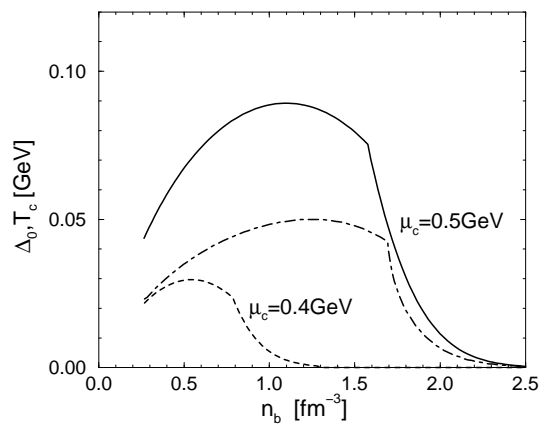}}
\caption{The gap $\Delta(\mu,T$=0) for $\mu_c$=0.4,~0.5~GeV (dashed and full
line, respectively) and critical temperature $T_c$ (dash\-dotted line,
$\mu_c$=0.5~GeV) versus baryon charge density $n_b$.  }
 \label{fig_gap}
\end{figure}
\noindent
Also plotted is the critical temperature $T_c$ at which superconductivity
disappears. The largest $T_c$ occurs at slightly higher densities than
the largest gap, but at fixed chemical potential the BCS relation $\Delta_0
=\pi T_c/\gamma$ approximately holds (where $\gamma\simeq$ 1.78 is Euler's
constant).

With $\mu$ exceeding the strange quark mass of $m_s^0\simeq$150~MeV, 
$s$-quarks are also present, which opens the possibility of 
forming new condensates. The instanton-induced pairing coupling constant 
in the $ud$ channel is ${\cal O}(m_s^0)$, while it is ${\cal O}(m_d^0)$  
(${\cal O}(m_u^0)$) for $us$ ($ds$). Therefore strange-light quark pairing 
is strongly suppressed,  presumably superseded by
perturbative Cou\-lomb interactions~\cite{BL}, 
which leads to $\Delta$$\simeq$1~MeV.
This  latter pairing involves $red$ (or color 3) light quarks left over
above, and it further breaks  
the remaining color subgroup $SU(2)_c$ 
by selecting, say,  scalar $d_3 s_{1}$, $u_3 s_2$ pairs.
What is finally left over in a ``normal'' state are $s_3$
quarks: according to~\cite{BL} their perturbative pairing may still occur
in a spin-1 channel~\cite{note_gap}.

5. Summarizing this work, we first recall the lessons learned from $N_c$=2 
QCD: here, diquarks are colorless baryons, and the scalar ones are
as different from all other baryons as pions are different from other 
mesons. Chiral symmetry is restored at a very small chemical potential
$\mu={\cal O}(m_q^{1/2})$, and the usual $\langle\bar qq\rangle$ condensate 
is replaced by a diquark $\langle qq\rangle$ condensate. Our simple
mean field approach, lattice and instanton simulations are consistent 
with each other.

  In the real world with $N_c$=3, scalar $ud$-diquarks are strongly 
bound. If not for confinement, a Bose condensed diquark gas would be 
the ground state at low densities. The transition from 
nuclear matter to diquark-condensed matter starts at densities 
 $n_b\simeq 0.7-1$ fm$^{-3}$ , as 
 a phase in which $both$ $\langle\bar q q\rangle$, 
$\langle q q\rangle$ are non-zero (it may, however, be unstable). 
An appreciable BCS-like gap $\Delta\simeq$ 100 MeV builds up towards  
 the chiral restoration point. 
Beyond, instantons become Debye-screened, so that a non-zero 
$\langle q q\rangle$ eventually only survives due to one-gluon exchange. 

 An interesting way to understand these phenomena is suggested by
comparing the features of the instanton ensemble at high $T$ and
high $\mu$. In both cases quark propagation in time direction is
favored over spacelike propagation (suppressed by $\exp(-\pi T r)$
and $\exp(i\mu r)$, respectively). As a result, the fermion
determinant leads to strong correlations among instantons.
Eventually, the  random ensemble breaks into small clusters
restoring chiral symmetry. These clusters are oriented in the
time direction: ``instanton-anti-instanton'' pairs at high $T$
and ``polymers'' at high density.

Acknowledgements:  Related work has been performed  
independently by M.~Alford, K.~Rajagopal and F.~Wilczek; 
we thank them for useful discussion. 
 Our work is partly supported by US DOE grants
DE-FG02-88ER40388 and DE-FG06-90ER40561. RR is also supported 
by the A.-v.-Humboldt foundation as a Feodor-Lynen fellow.

\end{narrowtext}


\begin{references}
\bibitem{DP_note} 
In~\cite{DP_diquarks} the number of Goldstone modes for $N_c$=$N_f$=2 
was (incorrectly) given as 9. The general analysis was done in the 
context of technicolor, see \cite{Peskin}. In our case, the pattern is
$SU(2N_f)\rightarrow Sp(2N_f)$.
\bibitem{note_collisions} 
We will see that color superconductivity requires high densities and
comparatively cool temperatures $T<T_c\simeq 50$ MeV, so that the 
creation of this phase in heavy-ion collisions appears doubtful.
\bibitem{note3} 
The scattering length reflects the diquark size, which we expect to be
on the order of the typical instanton radius $\bar\rho\simeq 1/3$ fm.
\end{references}
\end{document}